\documentclass[preprint,aps,showpacs,nofootinbib,preprintnumbers,amsmath,amssymb]{revtex4-1}
\usepackage{}
\usepackage{epsfig}
\usepackage{subfigure}
\usepackage{dcolumn}
\usepackage{bm}
\usepackage[usenames ,dvipsnames]{xcolor}
\usepackage{slashed}
\usepackage{graphicx,color}

\begin{document}
\title{Factorization and angular distribution asymmetries in charmful baryonic $B$ decays}
\author{Y.K. Hsiao and C.Q. Geng}
\affiliation{
Chongqing University of Posts \& Telecommunications, Chongqing, 400065, China\\
Physics Division, National Center for Theoretical Sciences, Hsinchu, Taiwan 300\\
Department of Physics, National Tsing Hua University, Hsinchu, Taiwan 300}
\date{\today}

\begin{abstract}
We examine the validity of the generalized factorization method and calculate
the angular correlations in the charmful three-body baryonic decays of $\bar B^0\to \Lambda \bar p D^{(*)+}$.
With the timelike baryonic form factors newly extracted from the measured  baryonic $B$ decays, we obtain 
${\cal B}(\bar B^0\to \Lambda \bar p D^+,\Lambda \bar p D^{*+})=
(1.85\pm 0.30,2.75\pm 0.24)\times 10^{-5}$ to agree with the recent data from the
BELLE Collaboration, which demonstrates that the  theoretical approach based on the factorization still works well.
For the angular distribution asymmetries, we find 
${\cal A}_{\theta}(\bar B^0\to \Lambda \bar p D^+,\Lambda\bar p D^{*+})=(-0.030\pm 0.002\,,+0.150\pm 0.000)$, 
which are consistent with the current measurements. Moreover, we predict that
${\cal A}_\theta(\bar B^0\to p\bar p D^{0},p\bar p D^{*0})=+0.04\pm 0.01$.
Future precise explorations of these angular correlations 
at BELLE and LHCb as well as super-BELLE 
are important to justify the present factorization approach 
in the charmful three-body baryonic decays.

\end{abstract}

\maketitle
\section{introduction}
Recently, the BELLE Collaboration has reported the branching ratios of 
$\bar B^0\to \Lambda \bar p D^{(*)+}$ along with the first angular distribution asymmetries
  measured in the charmful three-body baryonic $B\to {\bf B\bar B'}M_c$ decays,
given by~\cite{Chang:2015fja}
\begin{eqnarray}
\label{Data}
{\cal B}(\bar B^0\to \Lambda \bar p D^+)&=&
(25.1\pm 2.6\pm 3.5)\times 10^{-6}\,,\nonumber\\
{\cal B}(\bar B^0\to \Lambda \bar p D^{*+})&=&
(33.6\pm 6.3\pm 4.4)\times 10^{-6}\,,\nonumber\\
{\cal A}_\theta(\bar B^0\to \Lambda\bar p D^-)&=&-0.08\pm 0.10\,,\nonumber\\
{\cal A}_\theta(\bar B^0\to \Lambda\bar p D^{*-})&=&+0.55\pm 0.17\,,
\end{eqnarray}
with the subscript $\theta$  as the angle 
between $\bar p$ and $D^{(*)-}$ moving directions in the $\Lambda\bar p$ rest frame,
where ${\cal A}_\theta \equiv ({\cal B}_+ -{\cal B}_-)/({\cal B}_+ +{\cal B}_-)$ represents
the angular distribution asymmetry, with
${\cal B}_{+(-)}$ defined as the branching ratio of the positive (negative) cosine value.
The data in Eq.~(\ref{Data}) can be important  due to the fact that
$\bar B^0\to \Lambda \bar p D^{+}$ and $\Lambda \bar p D^{*+}$
are two of the few current-type processes among the richly observed baryonic $B$ decays,
connected to the timelike baryonic form factors 
via the vector and axial-vector quark currents.
Note that although
$\bar B^0\to \Lambda\bar p \pi^+$ and $B^-\to \Lambda\bar p \rho^0$ are related to
the timelike baryonic form factors, they also mix with the contributions from
the scalar and pseudoscalar currents via the penguin diagrams.

The decays of $\bar B^0\to \Lambda\bar p D^{(*)-}$ have been previously studied in Ref.~\cite{ppD} with the branching ratios predicted
to be $(3.4\pm0.2)\times 10^{-6}$ and $(11.9\pm2.7)\times 10^{-6}$, respectively,
which are obviously much lower than the current data in Eq.~(\ref{Data}) and regarded as the failure of the theoretical approach 
based on the factorization in Ref.~\cite{Chang:2015fja}. To resolve the problem, in this work we will evaluate the hadronic matrix elements
from  the observed baryonic $B$ decays directly instead of using the data of
$e^+ e^-\to p\bar p(n\bar n)$ ($p\bar p\to e^+ e^-$) in Ref.~\cite{ppD}.

Compared to the experimental result of ${\cal A}_\theta(\bar B^0\to \Lambda\bar p D^-)$ in Eq.~(\ref{Data}),
the measured value of 
${\cal A}_\theta(\bar B^0\to \Lambda\bar p \pi^-)=-0.41\pm 0.11\pm 0.03$~\cite{Wang:2007as}
as the charmless counterpart is unexpectedly large. 
Moreover, the experimental implication of
${\cal B}(\bar B^0\to \Lambda\bar p \pi^-)\sim 
{\cal B}(B^-\to \Lambda\bar p \pi^0)\sim 3\times 10^{-6}$~\cite{Wang:2007as} 
looks mysterious as it breaks the isospin symmetry.
Since the decays of $\bar B^0\to \Lambda\bar p D^{(*)-}$  simply proceed through
the (axial)vector currents from the tree contributions, one suspects that 
$|{\cal A}_\theta(\bar B^0\to \Lambda\bar p \pi^-)|\gg |{\cal A}_\theta(\bar B^0\to \Lambda\bar p D^-)|$
is due to the additional (pseudo)scalar currents from the penguin diagrams
in $\bar B^0\to \Lambda\bar p \pi^-$.
Likewise,  the charmless three-body baryonic decays of
$B^-\to p\bar p(\pi^-, K^-)$ receive
the main contributions from the tree and penguin diagrams, respectively,
which may result in the wrong sign of 
${\cal A}_\theta(B^-\to p\bar p\pi^-)\simeq -{\cal A}_\theta(B^-\to p\bar p K^-)$~\cite{a_theta,Aaij:2014tua}.
It is hence expected that $\bar B^0\to p\bar p D^{0}$ 
from the tree-level diagrams can be more associated with $B^-\to p\bar p \pi^-$.
Clearly, the systematic studies of the   
angular correlations in $B\to {\bf B\bar B'} M_c$ are needed.

Most importantly, since the theoretical approach 
for the three-body baryonic $B$ decays depends on the generalized factorization,
according to the comments in Ref.~\cite{Chang:2015fja},
if the calculations fail to explain the data,
it will indicate that the model parameters need to be revised and,
perhaps, some modification of the theoretical framework is required. 
Note that it is also commented in Ref.~\cite{Chang:2015fja} that
the factorization fails to provide a satisfactory explanation for
the $M$-$\bar p$ angular correlations in $B^-\to p\bar p K^-$, $B^0\to p\bar \Lambda \pi^-$ and 
$B\to p\bar p D$. 
However, it is clearly misleading as
${\cal A}_\theta(B^-\to p\bar p K^-)$ has been well studied in Ref.~\cite{Geng:2006wz}, whereas
${\cal A}_\theta(B\to p\bar p D)$ has  been neither measured experimentally nor  predicted theoretically.

In this report, we will study
$\bar B^0\to p\bar p D^{(*)0}$ and $\bar B^0\to \Lambda\bar p D^{(*)-}$
in order to approve the factorization approach. In addition,
we will calculate their angular distribution asymmetries to have the first theoretical predictions.
Moreover, some of these charmful asymmetries will be compared to
the charmless counterparts of  $B^-\to p\bar p K^-(\pi^-)$ and 
$\bar B^0\to \Lambda\bar p \pi^-$ ($B^-\to \Lambda\bar p \pi^0$).

\section{Formalism}
\begin{figure}[t!]
\centering
\includegraphics[width=2.5in]{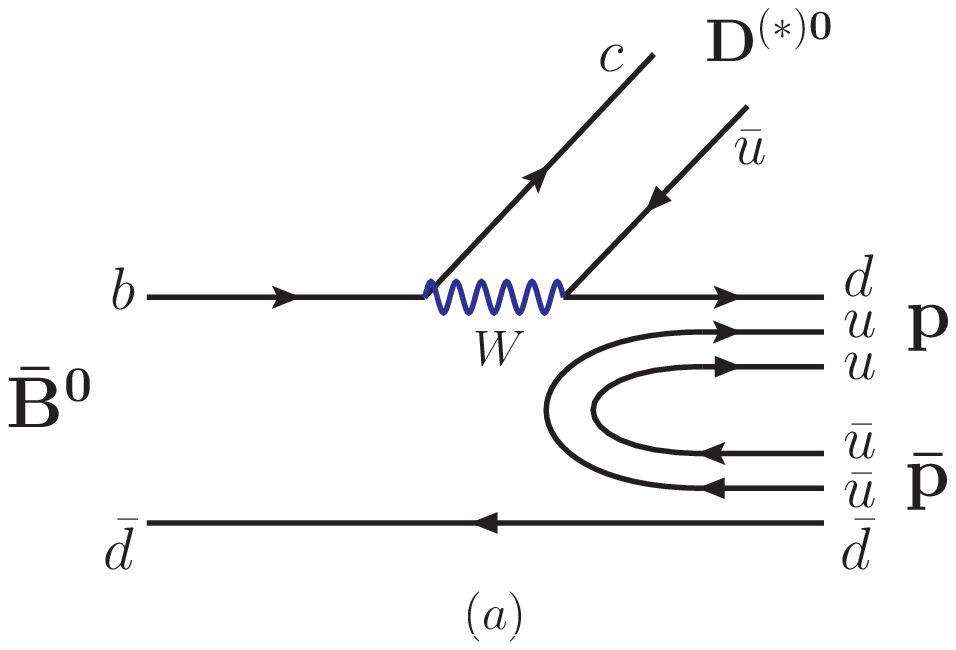}
\includegraphics[width=2.5in]{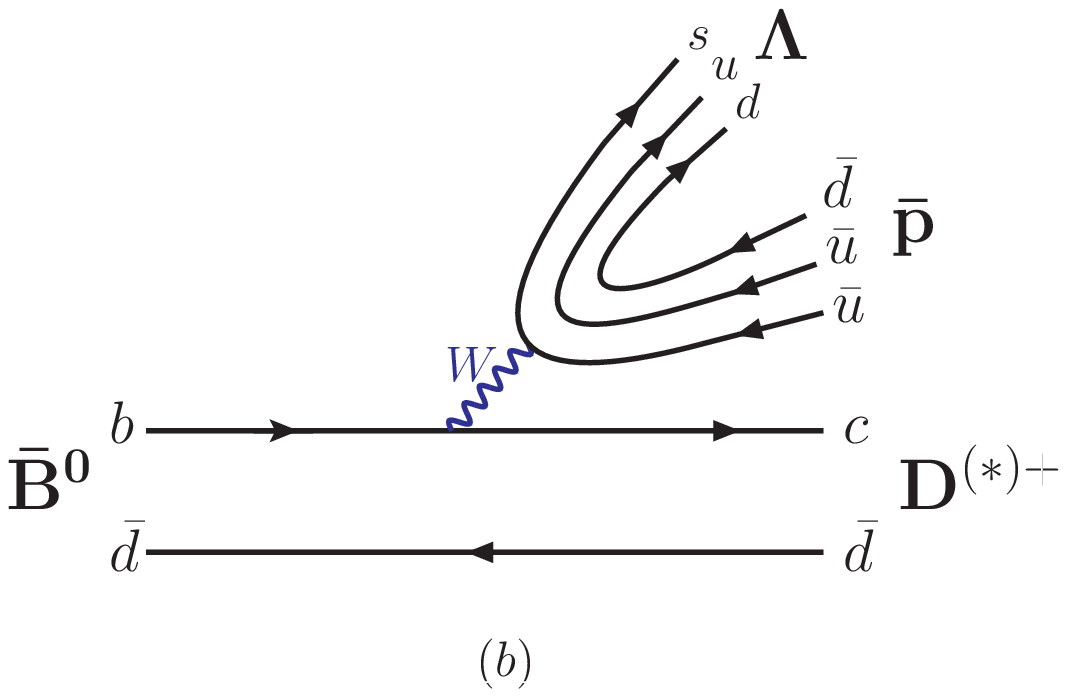}
\caption{Feynman diagrams for the three-body baryonic $B$ decays of 
(a)  $\bar B^0\to p\bar p D^{(*)0}$  and
(b)  $\bar B^0\to \Lambda\bar p D^{(*)+}$.}\label{dia}
\end{figure}
As shown in Fig.~\ref{dia}, in terms of the effective Hamiltonian for the quark-level $b\to c u\bar d(\bar s)$ transition
and the generalized factorization approach~\cite{ali}, 
the amplitudes of the $B\to {\bf B\bar B'}M_c$ decays can be written by~\cite{ppD}
\begin{eqnarray}\label{amp1}
&&{\cal A}(\bar B^0\to p\bar p D^{(*)0})=\frac{G_F}{\sqrt 2}V_{cb}V_{ud}^*
a_2\langle D^{(*)0}|(\bar c u)_{V-A}|0\rangle\langle p\bar p|(\bar d b)_{V-A}|\bar B^0\rangle\,,\nonumber\\
&&{\cal A}(\bar B^0\to \Lambda\bar p D^{(*)-})=\frac{G_F}{\sqrt 2}V_{cb}V_{us}^*
a_1\langle  \Lambda\bar p|(\bar s u)_{V-A}|0\rangle\langle D^{(*)-}|(\bar c b)_{V-A}|\bar B^0\rangle\,,
\end{eqnarray}
where $G_F$ is the Fermi constant, $V_{ij}$ are the CKM matrix elements,
 $(\bar q_1 q_2)_{V(A)}$ stands for $\bar q_1 \gamma_\mu(\gamma_5) q_2$, and
 $a_{1(2)}\equiv c^{eff}_{1(2)}+c^{eff}_{2(1)}/N_c^{eff}$ 
is composed of the effective Wilson coefficients $c_{1,2}^{eff}$ defined in Ref.~\cite{ali}.
In Eq.~(\ref{amp1}), the matrix elements  
for the $D^{(*)}$ meson productions through the $\bar c u$ quark currents can be written as
\begin{eqnarray}\label{dc}
\langle D|\bar c \gamma^\mu \gamma_5 u|0\rangle=-if_{D} p^\mu_D\;,
\langle D^{*}|\bar c \gamma^\mu u|0\rangle=m_{D^*}f_{D^*}\varepsilon^{\mu*}\;,
\end{eqnarray}
with $f_{D^{(*)}}$ the decay constant and $p_D^\mu$ ($\varepsilon^{\mu*}$)
the four-momentum (polarization). The matrix elements of 
the $B\to D^{(*)}$ transitions can be parametrized as~\cite{BSW}
\begin{eqnarray}\label{ff1}
\langle D| \bar c \gamma^\mu b|B\rangle&=&\bigg[(p_B+p_D)^\mu-\frac{m^2_B-m^2_D}{t}q^\mu\bigg]
F_1^{BD}(t)+\frac{m^2_B-m^2_D}{t}q^\mu F_0^{BD}(t)\,,\nonumber\\
\langle D^{*}|\bar c\gamma_\mu b|B\rangle&=&\epsilon_{\mu\nu\alpha\beta}
\varepsilon^{\ast\nu}p_B^{\alpha}p_{D^{*}}^{\beta}\frac{2V_1^{BD^*}(t)}{m_{B}+m_{D^{*}}}\;,\nonumber\\
\langle D^{*}|\bar c\gamma_\mu \gamma_5 b|B\rangle
&=&i\bigg[\varepsilon^\ast_\mu-\frac{\varepsilon^\ast\cdot q}{t}q_\mu\bigg](m_B+m_{D^{*}})A_1^{BD^*}(t)
 + i\frac{\varepsilon^\ast\cdot q}{t}q_\mu(2m_{D^{*}})A_0^{BD^*}(t)\nonumber\\
&-&i\bigg[(p_B+p_{D^{*}})_\mu-\frac{m^2_B-m^2_{D^{*}}}{t}q_\mu \bigg](\varepsilon^\ast\cdot q)\frac{A_2^{BD^*}(t)}{m_B+m_{D^{*}}}\;,
\end{eqnarray}
where $t\equiv q^2$ with $q=p_B-p_{D^{(*)}}=p_{\bf B}+p_{\bf\bar B'}$.
With the $\Lambda\bar p$ pair produced from the $s\bar u$ quark currents,
$\bar B^0\to \Lambda\bar p D^{(*)-}$ is classified as the current-type decay, 
such that
the matrix elements for the baryon pair production are in the forms of
\begin{eqnarray}\label{timelikeF}
\langle {\bf B}{\bf\bar B'}|\bar q_1\gamma_\mu q_2|0\rangle
&=&
\bar u\bigg\{F_1\gamma_\mu+\frac{F_2}{m_{\bf B}+m_{\bf \bar B'}}i\sigma_{\mu\nu}q_\mu\bigg\}v\;,\nonumber\\
&=& \bar u\bigg\{[F_1+F_2]\gamma_\mu+\frac{F_2}{m_{\bf B}+m_{\bf \bar B'}}(p_{\bf \bar B'}-p_{\bf B})_\mu\bigg\}v\;,\nonumber\\
\langle {\bf B}{\bf\bar B'}|\bar q_1\gamma_\mu \gamma_5 q_2|0\rangle
&=&\bar u\bigg\{g_A\gamma_\mu+\frac{h_A}{m_{\bf B}+m_{\bf \bar B'}}q_\mu\bigg\}\gamma_5 v\,,
\end{eqnarray}
where  $F_{1,2}$,
$g_A$ and $h_A$ are the timelike baryonic form factors, and
$u$($v$) is the (anti-)baryon spinor.
Being classified as the transition-type decays, the study of
$\bar B^0\to p\bar p D^{(*)0}$ needs to know the matrix elements for the $\bar B^0\to p\bar p$ transition,
which are parameterized as
\begin{eqnarray}\label{transitionF}
&&\langle {\bf B}{\bf\bar B'}|\bar q'\gamma_\mu b|B\rangle=
i\bar u[  g_1\gamma_{\mu}+g_2i\sigma_{\mu\nu}p^\nu +g_3 p_{\mu} 
+g_4q_\mu +g_5(p_{\bf\bar B'}-p_{\bf B})_\mu]\gamma_5v\,,\nonumber\\
&&\langle {\bf B}{\bf\bar B'}|\bar q'\gamma_\mu\gamma_5 b|B\rangle=
i\bar u[ f_1\gamma_{\mu}+f_2i\sigma_{\mu\nu}p^\nu +f_3 p_{\mu} 
+f_4q_\mu +f_5(p_{\bf\bar B'}-p_{\bf B})_\mu]v\,,
\end{eqnarray}
where $p=p_B-q$ and $g_i(f_i)$ $(i=1,2,3,4,5)$ are 
the $B\to{\bf B\bar B'}$ transition form factors.
The momentum dependences of the $B \to D^{(*)}$ transition form factors have been studied
in QCD models, given by~\cite{MFD}
\begin{eqnarray} \label{eq:FFpara1}
f(t)&=& {f(0)\over (1-t/M_{P(V)}^2)
[1-\sigma_1 t/M_{P(V)}^2+\sigma_2 t^2/M_{P(V)}^4]}\,,\;
\end{eqnarray}
for $f=F_1^{BD} (A_0^{BD^*}, V_1^{BD^*})$ and
\begin{eqnarray} \label{eq:FFpara2}
f(t) &=& {f(0)\over 1-\sigma_1 t/M_{V}^2+\sigma_2 t^2/M_{V}^4}\,,\;
\end{eqnarray}
for $f=F_0^{BD}$, $A_1^{BD^*}$ and $A_2^{BD^*}$,
while 
those of $F_1$ and $g_A$ in pQCD counting rules
can be written as \cite{Brodsky1,Brodsky2,Brodsky3}
\begin{eqnarray}\label{timelikeF2}
&&F_1=\frac{C_{F_1}}{t^2}\bigg[\text{ln}\bigg(\frac{t}{\Lambda_0^2}\bigg)\bigg]^{-\gamma}\;,
\qquad g_A=\frac{C_{g_A}}{t^2}\bigg[\text{ln}\bigg(\frac{t}{\Lambda_0^2}\bigg)\bigg]^{-\gamma}\;,
\end{eqnarray}
where $\gamma=2.148$
and $\Lambda_0=0.3$ GeV.
Note that $h_A=C_{h_A}/t^2$~\cite{Hsiao:2014zza}
is in accordance with the violated partial conservation of the axial-vector current,
whereas $F_2=F_1/(t\text{ln}[t/\Lambda_0^2])$~\cite{F2,F2b} 
is small to be safely neglected.
According to the principle of pQCD counting rules, 
one gluon to speed up the spectator quark within the $B$ meson
is required in the $B\to {\bf B\bar B'}$ transition, 
which causes an additional $1/t$ to $F_1$ and $g_A$, such that 
the momentum dependences of $f_i(g_i)$ can be written as~\cite{hou1}
\begin{eqnarray}\label{transitionF2}
f_i(t)=\frac{D_{f_i}}{t^3}\;, \qquad g_i(t)=\frac{D_{g_i}}{t^3}\;.
\end{eqnarray}
Furthermore, while the $SU(3)$ flavor symmetry can relate different decay modes,
the $SU(2)$ spin symmetry can combine
the vector and axialvector currents to be the chiral currents.
Consequently, one gets
the baryonic form factors 
to be~\cite{ppD,Hsiao:2014zza,Brodsky1,Brodsky2,Brodsky3,hou1,hou2}
\begin{eqnarray}\label{C&D}
&&C_{F_1}=C_{g_A}=-\sqrt\frac{3}{2}C_{||}\,,\;
C_{h_A}=-\frac{1}{\sqrt 6}(C_D+3C_F)\,,\nonumber\\
&&D_{g_1(f_1)}=\frac{1}{3}D_{||}\mp\frac{2}{3}D_{\overline{||}}\,,\;
D_{g_j(f_j)}=\mp\frac{1}{3}D_{||}^j\,,\nonumber\\
&&D_{g_1(f_1)}=-\sqrt {3\over 2}D_{||}\,,\;
D_{g_j(f_j)}=\mp\sqrt {3\over 2}D_{||}^j\,,
\end{eqnarray}
with the constants $C_{||}$, $C_{D(F)}$, $D_{||(\overline{||})}$, 
and $D_{||}^j$ ($j=2,3,4,5$) to be determined. 
Note that the relation for $C_{h_A}$ is simply from the $SU(3)$ symmetry.

To integrate over the phase space of the three-body $B\to{\bf B\bar B'}M_c$ decays,
we use~\cite{eqofag,Geng:2006wz}
\begin{eqnarray}\label{Gamma}
\Gamma=\int^{+1}_{-1}
\int^{(m_B-m_{M_c})^2}_{(m_{\bf B}+m_{\bf \bar B'})^2}
\frac{\beta_t^{1/2}\lambda^{1/2}_t}{(8\pi m_B)^3}|\bar {\cal A}|^2\;dt\;d\text{cos}\theta\;,
\end{eqnarray}
where  $\beta_t=1-(m_{\bf B}+m_{\bf \bar B'})^2/t$,
$\lambda_t=m_B^4+m_{M_c}^4+t^2-2m_{M_c}^2 t-2m_B^2 t-2m_{M_c}^2 m_B^2$,
the angle $\theta$ is between $\bf \bar B'$ and $M_c$ moving directions
in the $\bf B\bar B'$ rest frame, and 
$|\bar{\cal  A}|^2$ is the squared amplitude  of Eq.~(\ref{amp1})  by summing over all spins. 
Note that the $\bf B(\bar B')$ energy is given by
\begin{eqnarray}\label{Ep}
E_{\bf B(\bar B')}&=&
\frac{m_B^2+t-m_{\bf B(\bar B')}^2\mp \beta_t^{1/2}\lambda_t^{1/2}\cos\theta}{4m_B}\;.
\end{eqnarray}
From Eq. (\ref{Gamma}), we define the angular distribution asymmetry:
\begin{eqnarray}\label{AFB}
A_{\theta}\equiv\frac{\int^{+1}_0\frac{d\Gamma}{d\cos\theta}d\cos\theta
-\int^0_{-1}\frac{d\Gamma}{d\cos\theta}
d\cos\theta}{\int^{+1}_0\frac{d\Gamma}{d\cos\theta}
d\cos\theta+\int^0_{-1}\frac{d\Gamma}{d\cos\theta}
d\cos\theta}\;,
\end{eqnarray}
where $d\Gamma/d\cos\theta$ is a function of $\cos\theta$ known as the angular distribution,
which presents the $M_c$-$\bar B'$ angular correlation in  $B\to{\bf B\bar B'}M_c$.

\section{Numerical Analysis}

In our numerical analysis, the theoretical inputs of the CKM matrix elements in the
Wolfenstein parameterization  and the decay constants for $D^{(*)}$ 
are given by~\cite{pdg,Lucha:2014xla}
\begin{eqnarray}
&&(V_{cb},V_{ud},V_{us})=(A\lambda^2,1-\lambda^2/2, \lambda)\,,\nonumber\\
&&(\lambda,\,A,\,\rho,\,\eta)=(0.225,\,0.814,\,0.120\pm 0.022,\,0.362\pm 0.013)\,,\nonumber\\
&&(f_D,\,f_{D^*})=(204.6\pm 5.0,\;252.2\pm 22.7)\;\text{MeV}\,.
\end{eqnarray}
In Table~\ref{MF}, 
\begin{table}[!b]
\caption{ \sl The form factors of $B\to D^{(*)}$ at $t=0$
in Ref. \cite{MFD} with $M_P\simeq M_V=6.4$ GeV.
}\label{MF}
\begin{tabular}{|c|cccccc|}
\hline
$B\to D^{(*)}$&$F_1^{BD}$&$F_0^{BD}$&$V_1^{BD^*}$&$A_0^{BD^*}$&$A_1^{BD^*}$&$A_2^{BD^*}$\\\hline
f(0)                &0.67    &0.67   &0.76   &0.69    &0.66   &0.62\\
$\sigma_1$  &0.57     &0.78   &0.57   &0.58   &0.78   &1.40 \\
$\sigma_2$  &-----      & -----   &-----    &-----    &-----    &0.41 \\\hline
\end{tabular}
\end{table}
we adopt the $B\to D^{(*)}$ transition form factors 
from Ref.~\cite{MFD}, 
in which no uncertainty has been included. 
As mentioned early, 
the decays of $\bar B^0\to \Lambda \bar p D^{+}$ and $\bar B^0\to \Lambda \bar p D^{*+}$
belong to  the current-type modes, described by 
 the timelike baryonic form factors 
via the vector and axial-vector quark currents.
Note that 
$\bar B^0\to \Lambda\bar p \pi^+$ and $B^-\to \Lambda\bar p \rho^0$ are also connected to
the timelike baryonic form factors, but dominated by the additional ones via 
the scalar and pseudoscalar currents. 
With the extraction by the data from the current-type baryonic $B$ decays~\cite{Hsiao:2014zza},
 $F_1$ and $g_A$ as the timelike baryonic form factors can be given.
Because the $\bar B^0\to p\bar p$ 
transition form factors in $\bar B^0\to p\bar p D^{(*)0}$ 
are related to those of the charmless $B\to p\bar p M$ with $M=K^{(*)}$, $\pi(\rho)$
and the semileptonic $B^-\to p\bar p e^-\bar \nu_e$ decay, 
the extractions of $f_i(g_i)$ are also available~\cite{ppD}. It is hence determined that
\begin{eqnarray}\label{para}
&&(C_{||},\,C_D,\,C_F)=(111.4\pm 14.6,\,-6.8\pm 2.0,\,2.3\pm 0.9)\;{\rm GeV}^{4}\,,\nonumber\\
&&(D_{||},D_{\overline{||}})=(36.9\pm 45.9,-348.2\pm 18.7)\;{\rm GeV}^{5}\,,\nonumber\\ 
&&(D_{||}^2,D_{||}^3,D_{||}^4,D_{||}^5)=
(-44.7\pm 30.4,-426.7\pm 182.5, 4.3\pm 20.2,135.2\pm 29.4)\;{\rm GeV}^{4}\,.
\end{eqnarray}
In addition, $a_1$ and $a_2$ are fitted to be
\begin{eqnarray}
a_1=1.15\pm 0.04,\;a_2=0.40\pm 0.04\,.
\end{eqnarray}

As a result, 
we can reproduce the branching ratios shown in Table~\ref{table1}.
\begin{table}[b]
\caption{The data are from Refs.~\cite{pdg,Chang:2015fja,Chen:2011hy}.}\label{table1}
\begin{tabular}{|c|cc|}
\hline 
decay mode&data&our results\\\hline
$10^{4}{\cal B}(\bar B^0\to p\bar p D^{0})$            &$1.04\pm 0.07$&$1.04\pm 0.12$\\
$10^{4}{\cal B}(\bar B^0\to p\bar p D^{*0})$           &$0.99\pm 0.11$&$0.99\pm 0.09$\\
$10^{5}{\cal B}(\bar B^0\to \Lambda\bar p D^-)$    &$2.51\pm 0.44$&$1.85\pm 0.30$\\
$10^{5}{\cal B}(\bar B^0\to \Lambda\bar p D^{*-})$&$3.36\pm 0.77$&$2.75\pm 0.24$\\
\hline\hline
${\cal A}_\theta(\bar B^0\to p\bar p D^{0})$            &-----&$+0.04\pm 0.01$\\
${\cal A}_\theta(\bar B^0\to p\bar p D^{*0})$           &-----&$+0.04\pm 0.01$\\
${\cal A}_\theta(\bar B^0\to \Lambda\bar p D^-)$    &$-0.08\pm 0.10$&$-0.030\pm 0.002$\\
${\cal A}_\theta(\bar B^0\to \Lambda\bar p D^{*-})$&$+0.55\pm 0.17$&$+0.150\pm 0.000$\\
\hline
\end{tabular}
\end{table}
It should be pointed out that the main reason for 
the underestimated breaching ratios of $\bar B^0\to \Lambda\bar p D^{(*)-}$
in Ref.~\cite{ppD} is due to the small values of $F_1$ and $g_A$ extracted
from the data of $e^+ e^-\to p\bar p(n\bar n)$ ($p\bar p\to e^+ e^-$),
which are in fact related to the 
electromagnetic form factors of the proton (neutron) pair
without taking into account the timelike axial structures, 
induced from the weak currents due to $W$ and $Z$ bosons.
However, in this work, we take the data from the current-type baryonic $B$ decays as used in
 Ref.~\cite{Hsiao:2014zza}, which explains why
the data in Eq.~(\ref{Data}) of ${\cal B}(\bar B^0\to \Lambda\bar p D^{(*)-})$ can be explained.
With the current precise data for the axialvector current already,
 future new data should not change our present fitting parameters very much.

In the table,  we also show our predictions of the angular distribution asymmetries.
In particular, our result of ${\cal A}_\theta(\bar B^0\to \Lambda\bar p D^{-})=-0.030\pm0.002$ 
is consistent with the data in Eq.~(\ref{Data})~\cite{Chang:2015fja}, which shows that 
the unexpected large center number of ${\cal A}_\theta(\bar B^0\to \Lambda\bar p \pi^{-})=-30\%$
is either to be a much small value in the future measurement or 
 due to some unknown sources 
through the (pseudo)scalar currents from the penguin diagrams.
It is interesting to note that our prediction of ${\cal A}_\theta(\bar B^0\to \Lambda\bar p D^{*-})=0.150\pm 0.000$
is large but it is still lower than the data of $(55\pm 17)\%$ in Ref.~\cite{Chang:2015fja}.
Note that the small  uncertainty of our prediction
 results from the elimination of the timelike form factors by Eq.~(\ref{AFB}). 
The reason why the decay of $\bar B^0\to \Lambda\bar p D^{*-}$ can lead to 
a considerable large ${\cal A}_\theta\simeq 15\%$ is that,  
being one of the $B\to D^*$ transition form factors in Eq.~(\ref{ff1}),
the $V_1^{BD^*}$ term with $\epsilon_{\mu\nu\alpha\beta}$ is able to relate $F_1$ and $g_A$
from different currents,
such that $V_1^{BD^*} A_1^{BD^*} F_1 g_A (E_{\bar p}-E_p)$ can arise 
with $E_{\bar p}-E_p\propto \cos\theta$.  
It is important to point out that in the future experiments, our prediction of 
 ${\cal A}_\theta(\bar B^0\to p\bar p D^{0})=0.04\pm 0.01$ can be used to check
 if there is a simple relation between
$\bar B^0\to p\bar p D^{0}$ and $B^-\to p\bar p \pi^-$, which are both
dominated by the tree-level contributions. 
In addition, we remark that our
results are based on the form factors
in Table~\ref{MF} without any uncertainty included. If there are some possible errors,
our fitting values for the angular distributions could change.

\section{Conclusions}
We have revisited the charmful three-body baryonic  decays of $\bar B^0\to \Lambda \bar p D^{(*)+}$.
With the timelike baryonic form factors newly extracted from the brayonic $B$ decays instead of 
$e^+ e^-\to p\bar p(n\bar n)$ ($p\bar p\to e^+ e^-$), 
we have found that 
${\cal B}(\bar B^0\to \Lambda \bar p D^+,\Lambda \bar p D^{*+})=
(1.85\pm 0.30,2.75\pm 0.24)\times 10^{-5}$, which  agree with the data in Eq.~(\ref{Data}) from the 
BELLE Collaboration~\cite{Chang:2015fja}.
The agreement has demonstrated that our theoretical approach 
based on the factorization is still valid.
Clearly, the revision of model parameters and the modification of the factorization approach 
are not required unlike the statement in Ref.~\cite{Chang:2015fja}.

We have also studied the $M_c$-$\bf\bar B'$ angular distribution asymmetries in
the charmful baryonic $B$ decays of $B\to {\bf B\bar B'}M_c$.
Explicitly, we have obtained
${\cal A}_{\theta}(\bar B^0\to \Lambda \bar p D^+,\Lambda \bar p D^{*+})
=(-0.030\pm 0.002\,,+0.150\pm 0.000)$, which are  consistent with the current data.
In addition, we have predicted that
${\cal A}_\theta(\bar B^0\to p\bar p D^{0},p\bar p D^{*0})=+0.04\pm 0.01$. 
We believe that the future precision measurements  of ${\cal A}_\theta(B\to p\bar p D^{(*)},\Lambda\bar p D^{(*)})$
could be used to  compare with the charmless counterparts of ${\cal A}_\theta(B^-\to p\bar p K^-(\pi^-))$ and 
${\cal A}_\theta(B\to \Lambda\bar p \pi)$.
It is expected that the differences between the charmful and charmless cases, such as 
${\cal A}_\theta(\bar B^0\to \Lambda\bar p \pi^-)\simeq -41\%$ 
and ${\cal A}_\theta(\bar B^0\to \Lambda\bar p D^-)$, 
would be originated from different contributions at  tree and penguin levels.
Clearly, it is worthy to have close examinations of ${\cal A}_{\theta}(B\to {\bf B\bar B'}M_c)$ 
at BELLE and LHCb as well as the future super-B facilities.

\section*{ACKNOWLEDGMENTS}
The work was supported in part by National Center for Theoretical Science, National Sciences
Council (NSC-101-2112-M-007-006-MY3), MoST (MoST-104-2112-M-007-003-MY3) and National Tsing Hua
University (104N2724E1).

\end{document}